\documentclass[11pt,twoside]{article}
\usepackage{asp2014}

\aspSuppressVolSlug
\resetcounters

\begin{document}

\title{An Autograded, Open Educational Resource Assessment Tool for Astronomy Using MyOpenMath}
\author{Joshua Tan,$^1$ Allyson Sheffield,$^2$ and Jana Grcevich$^3$
\affil{$^1$LaGuardia Community College, City University of New York, Long Island City, New York, USA; \email{jotan@lagcc.cuny.edu}}
\affil{$^2$LaGuardia Community College, City University of New York, Long Island City, New York, USA; \email{asheffield@lagcc.cuny.edu}}
\affil{$^3$Columbia University, New York, New York, USA; \email{jgrcevich@astro.columbia.edu}}}

\paperauthor{Joshua Tan}{jotan@lagcc.cuny.edu}{0000-0003-3835-8115}{LaGuardia Community College, City University of New York}{Natural Sciences Department}{Long Island City}{New York}{11101}{USA}
\paperauthor{Allyson Sheffield}{asheffield@lagcc.cuny.edu}{0000-0003-2178-8792}{LaGuardia Community College, City University of New York}{Natural Sciences Department}{Long Island City}{New York}{11101}{USA}
\paperauthor{Jana Grcevich}{jgrcevich@astro.columbia.edu}{0000-0002-6521-1920}{Columbia University}{Astronomy Department}{New York}{New York}{10027}{USA}

\begin{abstract}
We present an Open Educational Resource system that instructors can use to assign automatically graded, randomized, and scaffolded assessment questions for astronomy classes. 
\end{abstract}

\section{Introduction}
In response to the ever-increasing prices of course texts and learning management systems (LMS), instructors in higher education have been adopting course materials that adhere to the Open Educational Resources (OER) model. First proposed during UNESCO's 2002 Forum on the Impact of Open Courseware for Higher Education in Developing Countries \citep{unesco2002OER}, the global OER movement has been in ascendancy since. Specifically designed to provide for educational resources under an open access model with a copy-left open license, the advantage of OER materials over traditional textbook and licensed instruments is that they are generally free, shareable, and modifiable. Under such models, most OER resources are made openly accessible online, the value of which became all the more apparent during the ongoing coronavirus pandemic when the majority of formal pedagogy moved to online platforms. Development of OER resources includes textbooks, stand-alone lesson plans and assessment tools, and even entire LMSs.

For teaching astronomy, a variety of OER materials are available, but the one with the most visibility and adopters is the OpenStax Astronomy textbook developed for introductory, non-major courses typified by the Astro 101 moniker \citep{fraknoi2019free}. This text is freely available through the OpenStax consortium\footnote{\url{https://openstax.org/details/books/astronomy}} and offers a competitive alternative to traditional textbooks from for-profit publishers. In addition to a 31-chapter textbook, supplemental materials include syllabi, answer guides, links to additional OER content, and a testbank of more than 1000 multiple choice questions available to instructors.

However, still underdeveloped are pedagogical resources for astronomy in the form of OER modules that can be connected to a course LMS with provisions for automated, randomized assessment. Most for-profit textbooks provide software modules at additional cost to students that can be customized by instructors including interactive simulations, virtual laboratories, and scaffolded assignments, but, to our knowledge, until this work there was no distributed, community-based OER alternative to these instruments.

For our implementation, we rely on the work started in 2005 by David Lippman, a mathematics professor at Pierce College in Washington State, USA. He is the primary author of a software package, IMathAS\footnote{\url{https://www.imathas.com/}}, which now serves as the backend for a free-to-use online LMS named MyOpenMath\footnote{\url{https://www.myopenmath.com/}}. The fully open-source IMathAS software is based on PHP\footnote{\url{https://www.php.net/}} and allows automated grading of student responses to assessment prompts. In addition to the classic LMS auto-graded question types of multiple choice and matching along with instructor-graded question types such as essays and file uploads, IMathAS allows instructors to write questions with answers that are numbers, strings, and even drawing objects with provisos for specified tolerances, alternate formats, and conditional feedback. Leveraging the capabilities of PHP libraries and hypertext editing, the questions can be written with dynamic content, links, embedded frames, tables, and plots. Randomizing macros, conditional feedback based on student answers, and scaffolded multi-part question types allow for questions that are uniquely tailored to individual students.

Aside from the question-writing capabilities, MyOpenMath allows instructors to create dynamical assignments with instructions and HTML-frames for content such as videos or applets. A flexible grading schema, individually adjustable timelines for due dates, and a practice mode provide additional functionality. Using Learning Tools Interoperability (LTI) packages and associated building blocks\footnote{\url{https://www.imsglobal.org/lti-fundamentals-faq}}, instructors can export individual assessments or entire courses into their home institution's LMS.

In the context of mathematics instruction, the technical features of the MyOpenMath system that allow for the individuation of assessments  were identified by \citet{macmahon2020mathematics} and \citet{hilton2013adoption} as innovating new mathematics assessment modalities and by \citet{sarmiento2019preventing} as an effective means to prevent academic dishonesty. MyOpenMath has also been shown to have positive outcomes in student experience and performance \citep{sarmiento2017student, sarmiento2018effect}.

\section{MyOpenMath Astronomy Library}

We undertook to create a public library of MyOpenMath questions and assessments that would take advantage of the benefits this system affords to astronomy instructors. The first questions entered the Astronomy Library in 2017, and it currently has 175 user-generated introductory level questions with various degrees of randomization. Additionally, 765 of the more than 1000 multiple choice test bank questions first authored by Andrew Fraknoi aligned with the OpenStax Astronomy text have been added to the library. As of writing, the most popular questions in the library have been used more in more than 40 different courses. Users of MyOpenMath with instructor accounts, granted for users who verify their faculty affiliation, can also search through the thousands of other questions in mathematics, physics, and chemistry and add questions to their own private library or to any of the existing public libraries.

Additionally, the platform can be used as a way to host virtual laboratory investigations by embedding various OER simulation applets including those hosted by PhET interactive simulations \citep{wieman2008phet}\footnote{\url{https://phet.colorado.edu/}} and Columbia Center for New Media Teaching and Learning's HTML5 ports\footnote{\url{https://ccnmtl.github.io/astro-simulations/}} of the popular University of Nebraska at Lincoln online astronomy applets \citep{lee2011using}. We also provide an Astronomy Laboratory Template course which includes assessments that each contain series of questions based on student interactions with these simulations. This entire course or individual assessments can be copied and modified with an instructor account. 

We invite the reader to experience a sample assessment hosted at the MyOpenMath site by starting a free account and enrolling in course 108874 (no enrollment key required) where a selection of questions are presented. Two of these questions are shown in Figure \ref{questions}. In addition, two sample questions from the Astronomy Laboratory Template course featuring the PhET Gravity and Orbits simulation\footnote{\url{https://phet.colorado.edu/en/simulation/gravity-and-orbits}} are demonstrated. For these two questions, hypothesis testing is scaffolded for the students who must choose a hypothesis before getting the parameters for a simulation to test their hypothesis. Credit is given not only for the correct measurement of the scenario but also for correctly identifying whether their initial hypothesis was correct or incorrect, thus modeling best practices of the scientific method.

\articlefiguretwo{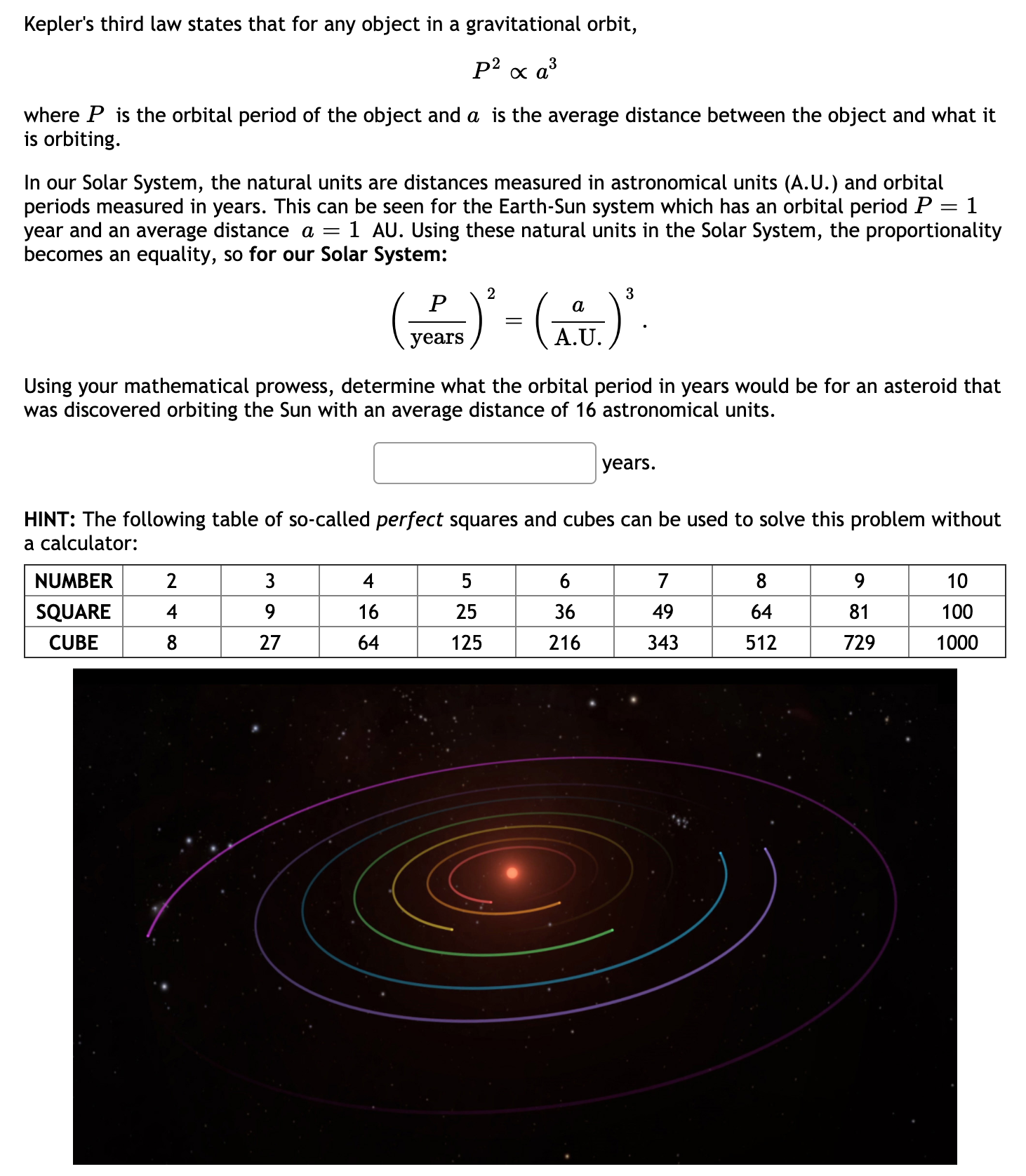}{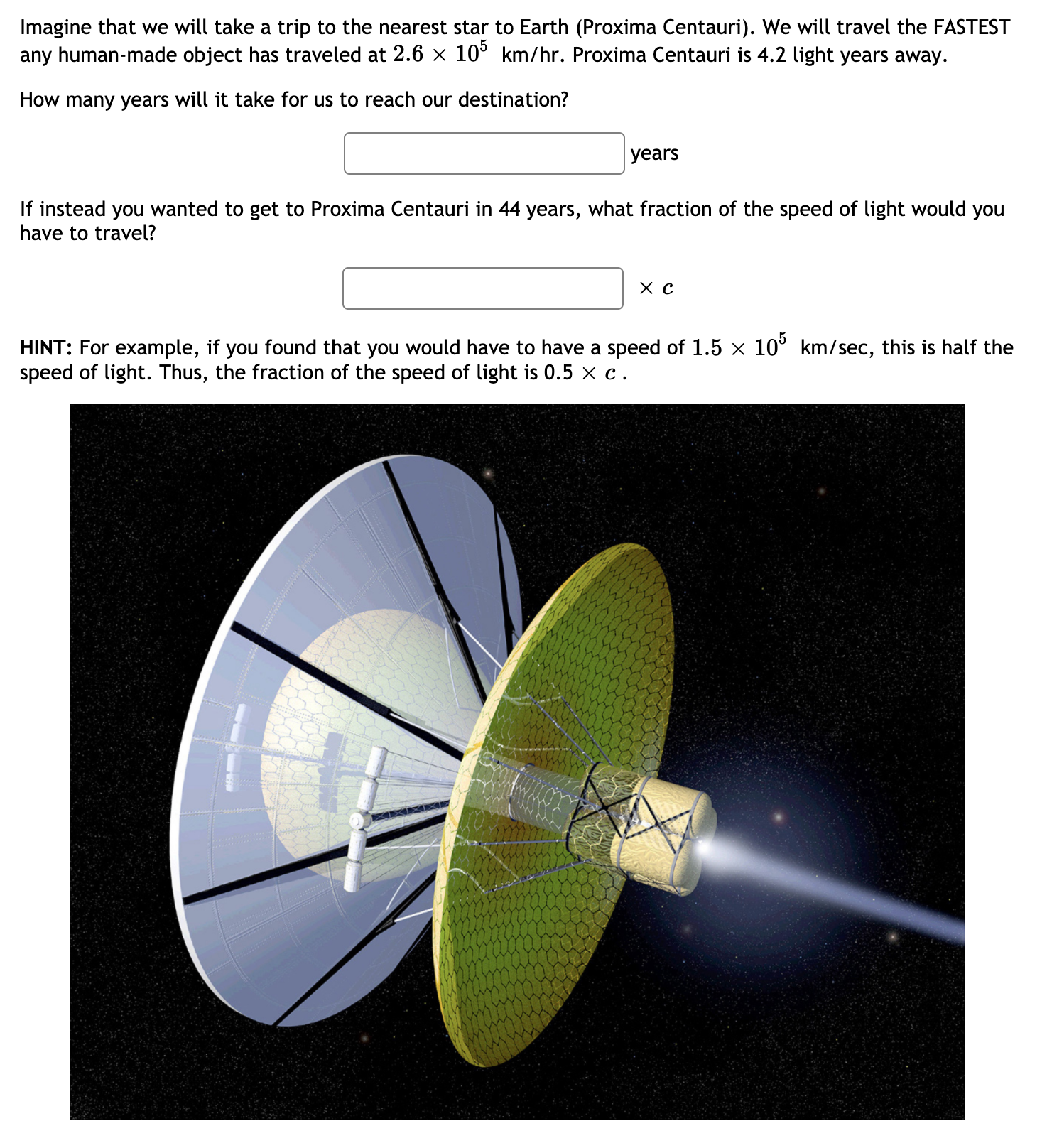}{questions}{Left, a question on Kepler's Third Law gives a random orbit from a list of nine possibilities. Since students may be asked to provide either the semi-major axis or the period of the orbit, there are eighteen different versions of this question. Image credit: NASA/JPL \url{https://www.spitzer.caltech.edu/explore/trappist-1} Right, a two-part question gives students a sense for the enormous distances between stars. The first part is the same for all instances of the question but the second has 40 different possible realizations randomized by student. Image credit: NASA \url{https://commons.wikimedia.org/wiki/File:Bussard_Interstellar_Ramjet_Engine.jpg}}
https://www.overleaf.com/project/6080818b66462145c3cf75c5
As a fully-realized OER system, this assessment regime can be utilized through MyOpenMath directly, or, if more control is desired, advanced users can download the IMathAS system along with the question libraries from MyOpenMath to host on their own server. As the software is released under GNU public license\footnote{\url{https://www.gnu.org/licenses/gpl-3.0.html}}, users are allowed to copy and modify the code as they wish. The Astronomy Library questions and assignment templates are additionally licensed under the Creative Commons CC-BY license\footnote{\url{https://creativecommons.org/licenses/}} which only requires simple attribution for further distribution of the material.

\section{Outcomes and Future Work}

In the four years since the first astronomy questions were written, more than 1000 students from City University of New York (CUNY)-LaGuardia Community College, the home institution of two of the authors, have used forms of these assessments through the MyOpenMath website or LTI integration into the CUNY Blackboard LMS. As part of this implementation together with the adoption of OpenStax Astronomy text, the course has been designated Zero Textbook Cost\footnote{\url{https://www.cuny.edu/libraries/open-educational-resources/guidelines-for-zero-textbook-cost-course-designation/}} which ensures that students who enroll in the course do not need to pay any additional amount for textbooks or homework modules. This was the first natural science course at LaGuardia to be so designated and over three years has saved students, the majority of whom come from low-income families, a combined estimated total of 100,000 USD. The access that a course fully using OER is meant to enable is thus realized in this context, and with more instructors from other institutions adopting the MyOpenMath astronomy modules, a growing collaboration and community of professionals dedicated to this ideal has the promise of increasing educational access much more widely.

We anticipate that future work will include not only expanding the question and assessment options available to instructors, but also the autograded question types. Work such as \citet{wiggins2019mixed} is ongoing to adopt machine learning techniques to provide autograded feedback for essay questions using the IMathAS system. Additional interactive laboratory investigations using real data and professional simulations are being developed and shared by the astronomical community through such instruments as Python\footnote{\url{https://www.python.org/}}-based platforms such as Jupyter Notebooks\footnote{\url{https://jupyter.org/install}} and Google Colab\footnote{\url{https://colab.research.google.com/}} \citep{newland2021remote}. Integrating these activities into IMathAS or MyOpenMath implementations would enable the extension of this model to more advanced astronomy classes and allow for authentic assessment experiences that mirror as closely as possible the typical workflow of an astronomer.

The possibilities are truly out of this world.

\acknowledgements Tan and Sheffield would like to acknowledge the support of the CUNY Office of Academic Affairs through a FY20 Open Educational Resources grant administered in part by LaGuardia Community College.

\bibliography{myopenmath} 

\end{document}